\documentclass[12pt]{article}
\usepackage[dvips]{graphicx}
\usepackage{pdproc}

\usepackage{graphicx}
\usepackage{dcolumn}
\usepackage{bm}
\usepackage{psfrag}
  \makeatletter 
  \def\@cite#1{[#1]} 
  \makeatother    
\begin{document}

\renewcommand{\thefootnote}{\alph{footnote}}

\title{
The Littlest Higgs Model and One-Loop Electroweak Precision Constraints
}

\author{MU-CHUN CHEN and SALLY DAWSON}

\address{ 
Physics Department, Brookhaven National Lab\\
Upton, New York, 11973, U.S.A.}

\abstract{
We present in this talk the one-loop electroweak precision constraints 
in the Littlest Higgs model, including the logarithmically enhanced 
contributions from both fermion and scalar loops. We find the one-loop 
contributions are comparable to the tree level corrections
in some regions of parameter space.   
A low cutoff scale is allowed for a non-zero triplet VEV. Constraints 
on various other parameters in the model are also discussed.
The role of triplet scalars in constructing a consistent 
renormalization scheme is emphasized.
\\
}

\normalsize\baselineskip=15pt

The Standard Model requires a Higgs boson to explain the generation
of fermion and gauge boson masses.  Precision electroweak measurements suggest
that the Higgs boson must be relatively light, $m_{H} <219~GeV$. 
Currently, experimental data overwhelmingly support the SM with 
a light Higgs boson. The simplest version of the Standard Model with a 
single Higgs boson, however, has the theoretical problem that 
the Higgs boson mass is quadratically sensitive 
to any new physics which may arise at high energy scales.
Little Higgs models are a new approach to understanding
the hierarchy between the $TeV$ scale of possible
new physics and the electroweak scale. These models have an expanded
gauge structure at the TeV scale which contains the Standard Model
$SU(2)\times U(1)$ electroweak gauge groups.  
The LH models are constructed
such that an approximate global symmetry prohibits the Higgs boson from 
obtaining a quadratically divergent mass until at least two loop order.
The Higgs boson is a pseudo-Goldstone 
boson resulting from the spontaneous breaking of the
approximate global symmetry and so
is naturally light. We present in this talk, which 
is based on the work done in Ref.~\cite{paper1}, 
the one-loop electroweak precision constraints 
in the Littlest Higgs model (LLH)~\cite{paper2}, 
which contains a gauged 
$[SU(2) \otimes U(1)]_{1} \otimes [SU(2) \otimes U(1)]_{2}$ 
symmetry as its subgroup. We  
include the logarithmically enhanced 
contributions from both fermion and scalar loops, and  emphasize 
the role of triplet scalars in constructing a consistent 
renormalization scheme. 

Precision electroweak measurements give stringent bounds on the scale of
little Higgs type models. One of the strongest bounds comes from fits
to the $\rho$ parameter, since in the LLH model the relation $\rho=1$ 
is modified  at the tree level. While the Standard Model requires 
three input parameters in the weak sector, a model with $\rho\ne 1$ 
at tree level, such as the LLH model, requires an additional input 
parameter in the gauge-fermion sector, 
which can be taken to be the 
VEV of the Higgs triplet, $v^\prime$. Many of the familiar predictions 
of the Standard Model are drastically changed by the need for an extra
input parameter~\cite{paper3,paper4}. We choose as our input parameters
the muon decay constant $G_{\mu}$, the physical Z-boson mass $M_{Z}^{2}$, 
the effective lepton mixing angle $s_{\theta}^{2}$ and the fine-structure 
constant $\alpha(M_{Z}^{2})$ as the four independent input parameters in 
the renormalization procedure. The $\rho$ parameter, defined as,
$\rho \equiv M_{W_{L}}^{2}/(M_{Z}^{2}c_{\theta}^{2})$, where $s_{\theta}^{2}$ 
is the effective leptonic mixing angle at the Z-resonance, 
and the $W$-boson mass, which is defined through muon decay, 
are then derived quantities.
Since the loop factor occurring in radiative corrections,  
$1/16\pi^2$, is similar in magnitude to the expansion parameter, 
$v^{2}/f^{2}$, of chiral perturbation 
theory, the one-loop radiative corrections can
be comparable in size to  the next-to-leading order contributions at tree 
level. We compute the loop corrections to the $\rho$ parameter which are
enhanced by large logarithms; we focus on terms of ${\cal O}\biggl(
{1\over 16\pi^2}\ln\biggl({M^2 \over Q^2}\biggr)\biggr)$, where 
$Q\sim M_Z$ and $M\sim f \sim {\cal O}(TeV)$.
At the one-loop level, we have to take into account the radiative correction 
to the muon decay constant $G_{\mu}$, the counterterm for the electric 
charge $e$, the mass counterterm of the Z-boson, and the counterterm 
for the leptonic mixing angle $s_{\theta}^{2}$. These corrections are collected  
in the quantity $\Delta r_{Z}$, 
\begin{equation}
s_{\theta}^{2}c_{\theta}^{2} = 
\frac{\pi \alpha(M_{Z}^{2})}{\sqrt{2}G_{\mu}M_{Z}^{2}\rho}
\biggl[1-\frac{\Delta s_{\theta}^{2}}{s_{\theta}^{2}}
+\frac{c^{2}s^{2}}{\sqrt{2}G_{\mu}f^{2}}
+\Delta r_{Z} \biggr],
\end{equation}
where
\begin{equation}\label{deltarz}
\Delta r_{Z} = -\frac{\delta G_{\mu}}{G_{\mu}} 
- \frac{\delta M_{Z}^{2}}{M_{Z}^{2}}
+ \frac{\delta \alpha}{\alpha}
- \biggl( \frac{c_{\theta}^{2}-s_{\theta}^{2}}{c_{\theta}^{2}} \biggr)
\frac{\delta s_{\theta}^{2}}{s_{\theta}^{2}} \;, 
\end{equation}
We note that $\Delta r_{Z}$ defined in Eq.(\ref{deltarz}) differs from 
the usual $\Delta \hat{r}_{Z}$ defined in the SM by an extra contribution 
due to the renormalization of $s_{\theta}^{2}$.
The counter terms are given by
\begin{eqnarray}
\frac{\delta G_{\mu}}{G_{\mu}} & = & -\frac{\Pi^{WW}(0)}{M_{W_{L}}^{2}} 
+ \delta_{V-B}
\\
\delta{M_{Z}^{2}} & = & Re\biggl(\Pi^{ZZ}(M_{Z}^{2})\biggr)
\\
\frac{\delta s_{\theta}^{2}}{s_{\theta}^{2}} & = & 
Re \biggl[ \; \biggl(\frac{c_{\theta}}{s_{\theta}} \biggr)\; \biggr[
\frac{\Pi^{\gamma Z}(M_{Z}^{2})}{M_{Z}^{2}}
+ \frac{v_{e}^{2}-a_{e}^{2}}{a_{e}} \Sigma_{A}^{e}(m_{e}^{2}) 
\\
& & \qquad \qquad 
- \frac{v_{e}}{2s_{\theta}c_{\theta}}
\biggl( \frac{\Lambda_{V}^{Z\overline{e}{e}}(M_{Z}^{2})}{v_{e}}
-\frac{\Lambda_{A}^{Z\overline{e}{e}}(M_{Z}^{2})}{a_{e}} \biggr) \;
\biggr]\;
\biggr]  \nonumber
\\
\frac{\delta \alpha}{\alpha} & = & \Pi^{\gamma \gamma \prime}(0) 
+ 2 (\frac{g_{V}^{e}-g_{A}^{e}}{Q_{e}}) \frac{\Pi^{\gamma Z}(0)}{M_{Z}^{2}}.
\end{eqnarray}
Solving for $M_{W_{L}}^{2}$ and $\rho$ iteratively, 
we obtain a prediction for the physical W-boson mass
\begin{equation}\label{eq:mw}
M_{W_{L}}^{2} = \frac{1}{2} \biggl[ a (1+\Delta \hat{r}) + 
\sqrt{a^{2}(1+\Delta \hat{r})^{2} + 4a\Pi^{WW}(0)} \biggr]
\end{equation}
where $a \equiv \pi \alpha(M_{Z}^{2}) / \sqrt{2}G_{\mu} s_{\theta}^{2}$, 
and $\Delta \hat{r}$ is defined as
\begin{equation} 
\Delta \hat{r} = \Delta r_{Z}-\frac{\Delta s_{\theta}^{2}}{s_{\theta}^{2}}
+\frac{c^{2}s^{2}}{\sqrt{2}G_{\mu}f^{2}}
-\frac{\Pi^{WW}(0)}{M_{W_{L}}^{2}} \; .
\end{equation}

We find that the one-loop contribution to $\Delta r_{Z}$ due to 
the SU(2) triplet scalar field, $\Phi$, scales as 
$1/[16\pi^{2} (1/v^{2}) (v^\prime/v)^{2} M_{\Phi}^{2}]$. 
In the limit $v^{'} = 0$ while keeping $f$ fixed, which is equivalent to turning 
off the coupling $\lambda_{h \Phi h}$ in the Coleman-Weinberg potential, 
the one loop contribution due to the SU(2) triplet, 
$\Delta r_{Z}^{s}$, vanishes. 
The large $f$ limit of the scalar one-loop contribution, 
$\Delta r_{Z}^{s}$, vanishes depending upon how the limit 
$f \rightarrow \infty$ is taken~\cite{paper1}.
As $f$ approaches infinity, the parameter $\mu^{2}$ (thus $v^{2}$) 
can be kept to be of the weak scale by fine-tuning the unknown coefficient 
in the mass term $\mu^{2}$  
in the Coleman-Weinberg potential  
while all dimensionless parameters remain of order one. 
The scalar one-loop contribution in this limit does {\it not} de-couple because 
$M_{\Phi}^{2}$ 
increases as $f^{2}$ which compensates the $1/f^{2}$ suppression from 
$v^{\prime 2} /v^{2}$. In this case, the SM Higgs mass $m_{H}$ is of the weak 
scale $v$. 
On the other hand, without the fine-tuning mentioned above, 
$v$ can be held constant while varying $f$, 
if the quartic coupling $\lambda_{h^{4}}$ (thus $\lambda_{\Phi^{2}}$) 
approaches infinity as $f^{2}/v^{2}$. 
This can be done by taking $a \sim f^{2}/v^{2}$ while keeping 
$a^{\prime}$ finite and $s$ and $s^{\prime}$ having specific values. 
The scalar one-loop contribution then scales as
\begin{equation}
\Delta r_{Z}^{s} 
\sim 
\frac{1}{v^{2}} (\frac{v^{\prime}}{v})^{2} M_{\Phi}^{2}
\sim 
(\frac{1}{v^{2}}) 
(\frac{\lambda_{h\Phi h}}{\lambda_{\Phi^{2}}})^{2} \frac{v^{2}}{f^{2}} 
\lambda_{\Phi^{2}} f^{2}
\rightarrow \frac{\lambda_{h\Phi h}^{2}}{\lambda_{\Phi^{2}}} \quad .
\end{equation}
Since the coupling constant $\lambda_{\Phi^{2}}$ must approach 
infinity in order to keep $v$ constant as we argue above, 
the scalar one-loop contribution 
$\Delta r_{Z}^{s}$ 
thus vanishes in the limit $f \rightarrow \infty$ with $v$ held fixed and
no fine tuning. 
In this case, $m_{H} \sim \mu$ scales with $f$.  

\begin{figure}[htb]
\begin{center}
\psfrag{M(theory)-M(exp) (GeV)}[][]{$M_{\mbox{theory}}
-M_{\mbox{exp}}$ (GeV)}
\psfrag{x_L}[][]{$x_{L}$}
\psfrag{Mw  }[][]{$\delta M_{W_{L}}$(total) $\qquad $}
\psfrag{Mz  }[][]{$\delta M_{Z}$(total) $\qquad $}
\psfrag{Mwtree  }[][]{$\delta M_{W_{L}}$(tree) $\quad $}
\psfrag{deltaMw}[][]{1 $\sigma$ limit on 
$\delta M_{W_{L}}$ (exp) $\qquad \qquad \qquad$ }
\psfrag{deltaMz}[][]{1 $\sigma$ limit on 
$\delta M_{Z}$ (exp) $\quad \qquad \qquad \qquad $}
\includegraphics*[angle=270,width=11cm]{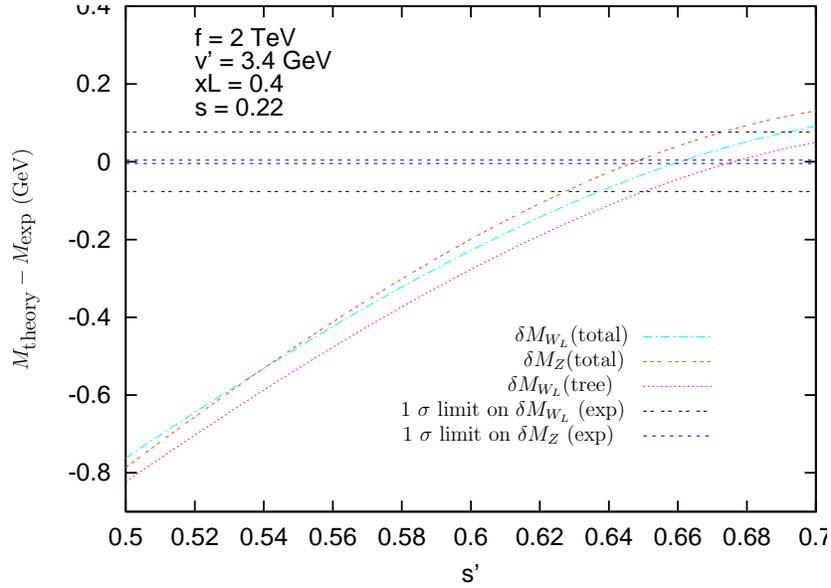}
\caption{%
Prediction for $M_{W_{L}}$ as a function of the mixing angle 
$s^\prime$ at the tree level and the one-loop level. 
Also plotted is the correlation between $M_{Z}$ and 
$s^\prime$ for fixed $s$, $v^{\prime}$ and $f$. 
The cutoff scale $f$ in this plot is $2$ $TeV$, the $SU(2)$ triplet VEV 
$v^\prime = 3.4 \; GeV$, the mixing angle $s=0.22$, and $x_{L}=0.4$.
}
\label{fig1}
\end{center}
\end{figure}

\begin{figure}[htb]
\begin{center}
\psfrag{xL = 0.4}[][]{\large $\large x_{L} = 0.4$}
\psfrag{s ' = 0.5}[][]{\large $\large s^{\prime} = 0.5$}
\psfrag{s = 0.2}[][]{\large $\large s = 0.2$}
\psfrag{v ' = 1 GeV}[][]{\large $\large v^{\prime} = 1 GeV$}
\psfrag{Dtree  }[][]{\large $\large \Delta_{\mbox{tree}}\qquad \qquad $  }
\psfrag{Df  }[][]{\large $\large \Delta r_{Z}^{f} \qquad \qquad \qquad \; $}
\psfrag{Ds  }[][]{\large $\large \Delta r_{Z}^{S} \qquad \qquad \qquad \; $}
\psfrag{D1loop  }[][]{\large $\large \Delta \hat{r}_{Z} -\Delta_{\mbox{tree}} \qquad   $}
\psfrag{piww  }[][]{\large $\large \Pi^{WW}(0)/M_{Z}^{2}  \qquad $}
\includegraphics*[angle=270,width=11cm]{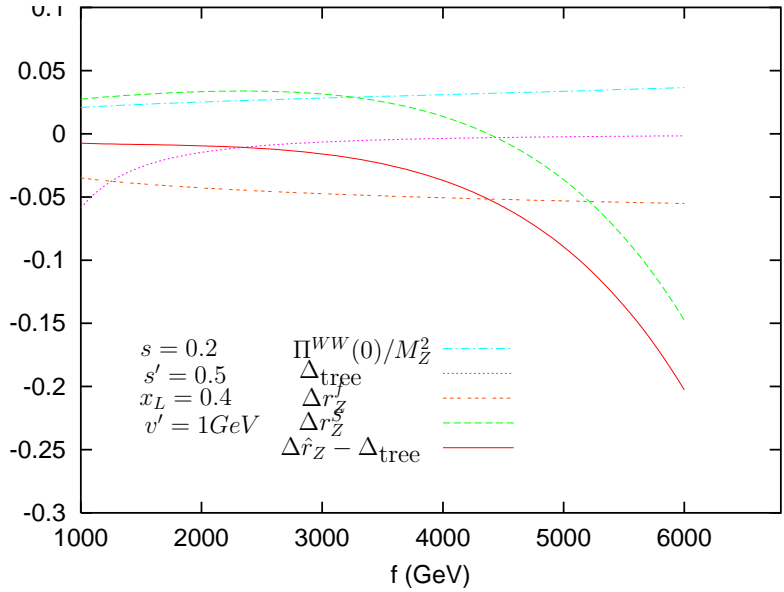}
\caption{%
The tree level correction, $\Delta_{\mbox{\tiny tree}}$, the 
fermionic and scalar contributions to the one loop correction, 
$\Delta r_{Z}^{f} $  
and $\Delta r_{Z}^{S}$,
the total one loop correction, $\Delta \hat{r}-\Delta_{\mbox{tree}}$, 
and $\Pi^{WW}(0)/M_{Z}^2$ 
as functions of the cutoff scale $f$ at fixed $s$, $s^{\prime}$, 
$x_{L}$ and $v^{\prime}$.
}
\label{fig2}
\end{center}
\end{figure}

We analyze the dependence of the W-boson mass, $M_{W_{L}}$, on the 
mixing between 
$SU(2)_{1}$ and $SU(2)_{2}$, described by $s^{\prime}$, 
the mixing between $U(1)_{1}$ 
and $U(1)_{2}$, described by $s$, the mixing parameter in $t-T$ 
sector, $x_{L}$, 
and the VEV of the $SU(2)$, 
$v^{\prime}$. The predictions for $M_{W_{L}}$ with and without the 
one-loop contributions for $f=2$ TeV is given in Fig.~\ref{fig1}, 
which demonstrates that a low value of $f$ ($f\sim 2~TeV$) is allowed by 
the experimental restrictions from the $W$ and $Z$ boson masses, 
provided the VEV of the $SU(2)$ triplet scalar field is non-zero. 
This shows the importance of the $SU(2)$ triplet in placing the electroweak 
precision constraints. In order to have experimentally acceptable 
gauge boson masses, however, 
the parameters of the model must be quite finely tuned, regardless 
of the value of the scale $f$. 
On the other hand, the prediction for $M_{W_{L}}$ is very 
sensitive to the values of $s^{\prime}$ as well as $v^{\prime}$. 
The non-decoupling of the SU(2) triplet scalar field shown in 
Fig.~\ref{fig2} implies the 
importance of the inclusion of the scalar one-loop 
contributions in the analyses.
In the region below $f=4~TeV$, where the tree level corrections are large, 
the vector boson self-energy is about half of the size of the tree 
level contributions, but 
with an opposite sign. (Other one-loop contributions roughly 
cancel among themselves in this region). Due to this cancellation 
between the tree level correction and the one-loop
correction, there is an allowed region of  parameter space
with low cutoff scale $f$.
Fig.~\ref{fig2} also shows that the tree level contribution of the LH
model get smaller as $f$ increases, as is expected.
In order to be consistent with experimental data, the triplet 
VEV $v^{\prime}$ 
must approach zero as $f$ goes to infinity.

Our results emphasize the need for a full one loop calculation.
\bibliographystyle{plain}

\end{document}